\documentclass{article}

\usepackage{arxiv}

\usepackage[utf8]{inputenc} 
\usepackage[T1]{fontenc}    
\usepackage{hyperref}       
\usepackage{url}            
\usepackage{booktabs}       
\usepackage{amsfonts}       
\usepackage{nicefrac}       
\usepackage{microtype}      
\usepackage{lipsum}

\usepackage{cite}
\usepackage[inline]{enumitem}

\usepackage{cuted}

\usepackage[numbers,sort]{natbib}

\usepackage{graphicx}
\usepackage[justification=centering]{caption}
\usepackage{wrapfig}
\usepackage{subcaption}

\usepackage{color}

\usepackage{amsmath}

\title{An Improved Quadrature Voltage-Controlled Oscillator with Through-Silicon-Via Inductor in Three-dimensional Integrated Circuits}

\author{
  Dawei Li \\
  Department of Electrical  and  Information Engineering\\
  South Central University for Nationalities\\
  Wuhan, China 430074 \\
  \texttt{leedavidhust@outlook.com} \\
   \And
 Madhava Sarma Vemuri 
 \\
  Department of Electrical and Computer Engineering\\
  North Dakota State University\\
  Fargo, North Dakota 58105 \\
  \texttt{madhava.vemuri@ndsu.edu} \\
  \AND
   Umamaheswara Rao Tida \\
  Department of Electrical and Computer Engineering\\
  North Dakota State University\\
  Fargo, North Dakota 58105\\
  \texttt{umamaheswara.tida@ndsu.edu} \\
}

\begin{document}
\maketitle

\begin{abstract}
Low-power quadrature voltage-controlled oscillator (QVCO) design utilizing transformer-feedback and current-reuse techniques with increased frequency range is proposed in this paper. With increasing demand for QVCOs in on-chip applications, the conventional spiral inductor based approaches for QVCOs has become a major bottleneck due to their large size. To address this concern, we propose to replace the conventional spiral inductor based approaches with through-silicon-via (TSV) inductor based approach in three-dimensional integrated circuits (3D ICs). In addition, the proposed QVCO circuit can provide higher frequency range of operation compared with conventional designs. Experimental results show  by replacing conventional spiral transformers with TSV transformers, up to 3.9x reduction in metal resource consumption. The proposed QVCOs achieves a phase noise of -114 $dBc/Hz$@1 $MHz$ and -111.2 $dBc/Hz$@1 $MHz$ at the carrier of 2.5 $GHz$ for toroidal TSV transformed based-QVCO and vertical spiral transformer based-QVCO respectively. The power consumption is only 1.5 $mW$ and 1.7 $mW$ for toroidal TSV transformed based-QVCO and vertical spiral transformer based-QVCO respectively, under the supply voltage of 0.7 $V$. 
\end{abstract}


\section{Introduction}

Voltage-controlled oscillator (VCO) is an essential component and extensively used for various on-chip applications i.e., phase-locked loops (PLLs) \cite{liu2017temperature, you201412ghz, ikeda20170}, clock distribution networks, signal generators, function generators, communication networks \cite{zhang2014design} etc. For all these applications, a VCO with low-power, low-noise, and low-resource consumption is desired because the VCO usually works at higher frequencies, and is the most power-hungry block as the power dissipation is proportional to the operating frequency \cite{zhang2014design,ikeda20170}. Hence, an efficient VCO design for these applications is required. 

VCOs can be implemented in two ways: \begin{enumerate*} \item Active VCO and \item{Passive VCO}\end{enumerate*}. Active VCOs do not contain passive inductor and hence these VCOs can be compact on-chip \cite{li2019optimal}. However, they do not perform well at higher frequencies due to high jitter associated with the transistors. The active VCOs also suffer from a severe frequency drift caused by the parasitic capacitors and resistors. Passive VCO works well at higher frequencies which most applications need and also has low jitter compared with  its active couterpart due to the reduced noise. Therefore, passive VCOs are more practical and attractive for various applications. In the rest of the paper, VCO refers to the passive VCO unless specified.

Inductors required for VCO can be implemented in various approaches: \begin{enumerate*}\item Off-chip Inductor, \item On-package inductor and \item On-chip inductor \end{enumerate*}. Off-chip inductor is bulky and is not practical for condensed integration \cite{yu2018batteryless,wang2015analytical}. On-package inductor suffers from poor scalability \cite{tan2014design}. Therefore, current research focuses on implementing on-chip inductors effectively for various applications.

Conventional on-chip spiral inductors occupy huge metal routing resources which limits the number of inductors that can be implemented on-chip. The following works discusses about the implementation of conventional on-chip inductors and their metal resource consumption: 
\begin{itemize} 
\item On-chip DC-DC converters with a conventional spiral inductor designed in \cite{wang2010integrated} requires $78,400 \mu m^2$ of metal area, equivalent to the area of 62K gates in 45 $nm$ technology. 
\item A 5 $GHz$ MIMO transceiver with 20 on-chip inductors implemented in \cite{palaskas20065} occupies a huge area of $18 \ mm^2$. 
\item A variable bandwidth 3-4 GHz transceiver for Internet of Things (IoT)  applications is proposed in \cite{liu20170} has 10 optimized  on-chip circular shape spiral inductors with an area of $5.6 \ mm^2$.
\end{itemize}

These on-chip planar inductors usually occupy large area, contributing to an increasing manufacture cost which is one of the main concerns for future ubiquitous applications. Therefore, compact on-chip inductors are needed for effective implementation of on-chip inductors for various applications.

On the other hand, three-dimensional integrated circuits (3D ICs) have become a promising alternative to keep Moore's law since an
 extra dimension makes chip size smaller. This technology has gained popularity for high computing devices like large scale processors as well as for compact devices like mobile chips. While there are many challenges still exist in 3D IC practicality, a major one is related to TSVs because they are huge and they do not scale with logic gates. In addition, lot of dummy TSVs are needed to be inserted to satisfy minimum density rule due to chemical mechanical polishing purposes. For example, Tezzaron needs one TSV for every 250 $\mu m\ \times$ 250 $\mu m$ area \cite{lim20133d}. This further increases the overhead by TSVs. One way to address this problem is to make on-chip devices using TSVs and utilize them in on-chip applications.

In our previous works, we addressed the on-chip inductor metal resource consumption by replacing the conventional spiral inductor with TSV inductor  for on-chip regulator and resonant clocking applications \cite{tida2014through, tida2014efficacy, tida2014green, tida2018dynamic,  tida2014opportunistic, tida2019single, kankonkar2016pwm, tida2014novel}. For on-chip regulator applications \cite{tida2014novel,kankonkar2016pwm, tida2014green, tida2019single}, the inductor area reduced by upto 4.3x by replacing spiral inductor with TSV inductor for the same performance of the regulator. For resonant clocking applications \cite{tida2014opportunistic, tida2014green, tida2018dynamic}, the inductor area reduced by upto 6.3x by replacing spiral inductor with TSV inductor for the same power reduction. Also, micro-channel shielding technique \cite{tida2014through, tida2014efficacy} is proposed to make TSV inductors practical at high frequencies. For these applications, TSV inductors are effective in reducing the metal resource consumption. 

VCO is another important component that is required for various on-chip applications. Conventionally on-chip VCO is designed by utilizing conventional spiral inductor which may not be efficient in utilizing the metal routing resources effectively. Towards this, we propose to use TSV inductors with novel coupling structures to enable efficient implementation of VCO. With these objectives, the main contributions of this work are as follows:
\begin{enumerate}
    \item Quantitative analysis of the Quadrature voltage controlled oscillator (QVCO) is discussed in detail.
    \item We propose to use novel transformer structures using TSV inductors to design  QVCO. Compared with conventional spiral implementation, TSV inductor based transformers will better utilize the metal routing resources.
    \item We design a low-supply QVCO with proposed TSV transformers. To author’s best knowledge, this paper is the first to design a TSV-based QVCO under a 0.7V supply. 
\end{enumerate}

The remainder of the paper is organized as follows: Section \ref{sec:background} reviews the background on VCO and TSV inductor design. A detailed description of QVCO design and the importance of transformer design on the performance of QVCO is given in Section \ref{sec:QVCO}. Novel TSV inductor structures to implement on-chip transformer for the design of QVCO is discussed in Section \ref{sec:TSVIndDes}. Simulation results of this QVCO and the comparison with other works is presented in Section \ref{sec:simResults} and concluding remarks are given in Section \ref{sec:conclusions}.

\section{Preliminaries} \label{sec:background}

\subsection{Basic VCO}

VCO generates sinusoidal signals whose frequency is varied with the input voltage of the oscillator. This circuit is extensively used in the design of integrated circuits (ICs) for various applications that include function generators, clock generators, phase-locked loops, wireless transceivers etc. VCO is usually power hungry due to its higher frequency of operation and hence an efficient VCO design is required.

\begin{figure}[!ht] 
  \centering
  \includegraphics[scale=0.6]{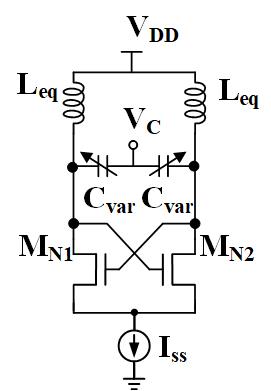}
  \caption{Circuit diagram of a NMOS based LC-VCO}
   \label{fig:basicOsc}
\end{figure}

The circuit diagram of a basic VCO is shown in Figure \ref{fig:basicOsc}. This circuit has three components: \begin{enumerate*} \item LC tank, \item NMOS cross-coupled pair and, \item Constant current source \end{enumerate*}. This circuit generates oscillations with the resonance frequency given by the equation \ref{eq:oscFreq} which is determined by the LC tank of this circuit when it is properly designed. Here $L_{eq}$ and $C_{eq}$ are the equivalent inductance and capacitance of the LC tank considering the parasitics of these components when implemented on-chip. The NMOS cross-coupled pair helps to achieve the phase shift of $360^o$ across the loop at the resonance frequency and compensates for the series resistance of the inductor for the oscillations to sustain. Current source helps to lower the supply voltage sensitivity for stable operation. For the circuit to oscillate at the resonance frequency, the loop gain should be greater than 1 and can be achieved when $R_L\geq \frac{2}{g_m}$ where $R_L$ is the equivalent series resistance of the inductor and $g_m$ is the transconductance of the NMOS transistor in the cross-coupled pair. Therefore, with proper sizing of the cross-coupled pair transistors and the input bias, this circuit generates oscillations with resonance frequency. For the VCO, the oscillation frequency is controlled by the input voltage $V_c$. This property can be achieved by implementing a variable MOS capacitor with the input voltage. 

\begin{equation}\label{eq:oscFreq}
\omega_{osc}=\frac{1}{\sqrt{L_{eq}C_{eq}}}
\end{equation}

Another important aspect with the design of VCO is to meet the specifications for the given application. The important specifications correspond to the VCO design are center frequency ($f_c$), tuning frequency range ($f_{min}$ to $f_{max}$), VCO gain $K_{VCO}$ and phase noise. Center frequency $f_c$ and tuning frequency range ($f_{min}$ to $f_{max}$) are determined by the frequency standards for the protocol environment like mobile communications, PLL design etc. VCO gain $K_{VCO}$ is determined by the control voltage sensitivity of the VCO and should be as small as possible in order to avoid stability issues. Phase noise $\phi_{noise}$ is another important specification which is a measure of signal quality in the frequency domain that determines the frequency fluctuations in a signal and is usually measured at 1 $MHz$.


Since most of the current wireless communication systems like LTE and WiMAX \cite{arai2014self} are employing quadrature modulation for up- and down-conversion, Quadrature VCOs (QVCOs)  are developed by coupling two differential VCOs . 
Therefore, QVCO requires  additional inductors  and  thus demands  more  metal resources  for  their  on-chip  implementation.  Towards  this,  in this   work  we  propose  to  use  TSV  inductor  based  QVCO  for compact design with better efficiency.


\subsection{On-chip Inductors}
The major bottleneck for many applications that need inductors to fully integrate on-chip is the metal resource consumption especially by the conventional spiral inductor technique. Hence it is crucial to have a high-inductance density on-chip inductor with the desired quality factor Q. In our previous works, we tackled this challenge by designing TSV inductor that performs on-par with the conventional spiral inductor for on-chip regulator and resonant clock distribution network applications with the metal resource consumption reduced by upto 6x, which is a huge benefit for realizing more on-chip applications that need inductors.

\begin{figure}[!ht] 
  \centering
  \includegraphics[scale=1]{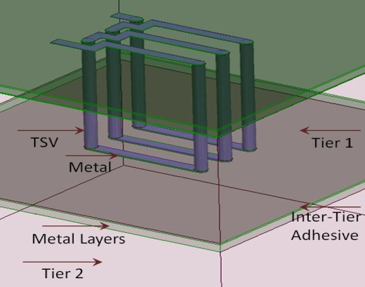}
  \caption{Three-turn toroidal TSV inductor \cite{tida2014novel}}
   \label{fig:simpleInd}
\end{figure}

A simple three-turn toroidal TSV inductor structure is shown in Figure \ref{fig:simpleInd}. From the figure, we can see that the TSV inductor is buried in the substrate and hence the substrate losses are higher compared with the conventional spiral inductor.  Our previous work \cite{tida2014efficacy} shows that these losses are not significant for low-frequency range and can be reduced by using micro-channel etching technique at high frequencies to make them practical.

As discussed earlier, on-chip QVCO implementation demands huge metal resources and has become a tradeoff for many applications that require QVCO especially for LTE, WiMAX etc. In order to make QVCO designs more practical for on-chip implementation in 3D technology, we propose to utilize TSV inductor structures that require less metal resources compared with the conventional spiral inductor structures.

\section{Transformer-feedback Current-reuse QVCO}\label{sec:QVCO}

In this section, we discuss about the design and analysis of QVCO for low power applications by combing two techniques: \begin{enumerate*}\item Transformer feedback and, \item Current reuse \end{enumerate*}. We first discuss briefly about these techniques individually on the two-phase VCO and then we present a circuit topology for low-power QVCO that combines both these techniques.

\begin{figure}[!ht] 
  \centering
   \includegraphics[scale=0.6]{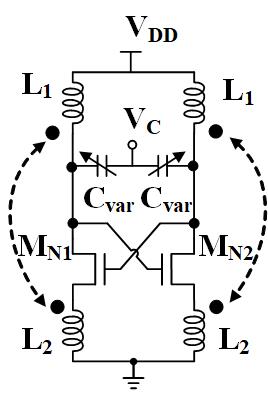}
  \caption{Circuit diagram of a TF-VCO}
   \label{fig:TFOsc}
\end{figure}

The simple differential VCO circuit shown in Figure \ref{fig:basicOsc} cannot be used for the low supply voltage operation due to the limited voltage headroom by the tail current source and the active circuits. The simple differential VCO circuit can be modified by utilizing transformer as shown in the Figure \ref{fig:TFOsc} and this modifed circuit is called two-phase transformer-feedback  VCO (TF-VCO). With the two-phase TF-VCO , the output voltage can swing above the supply voltage $V_{dd}$ and below the ground potential. Also, the drain and source signals are in phase for this TF-VCO circuit \cite{kwok2005ultra}. Therefore, the oscillation amplitude range for this circuit is increased for the given supply voltage. In other words, we can reduce the supply voltage for the given oscillator specifications i.e., phase noise and oscillation amplitude and hence the power consumption can be reduced. Another advantage with this technique is the improved voltage sensitivity due to the transformer and hence the tail current supply is not required.

\begin{figure}[!ht] 
  \centering
  \includegraphics[scale=0.6]{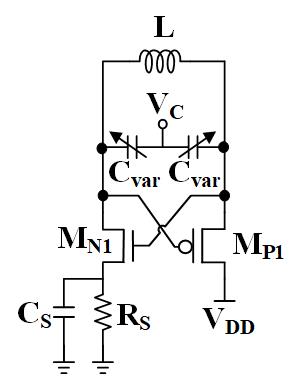}
  \caption{Circuit diagram of a CR-VCO}
   \label{fig:CROsc}
\end{figure}

\begin{figure*}[!ht] 
  \centering
  \includegraphics[scale=0.45]{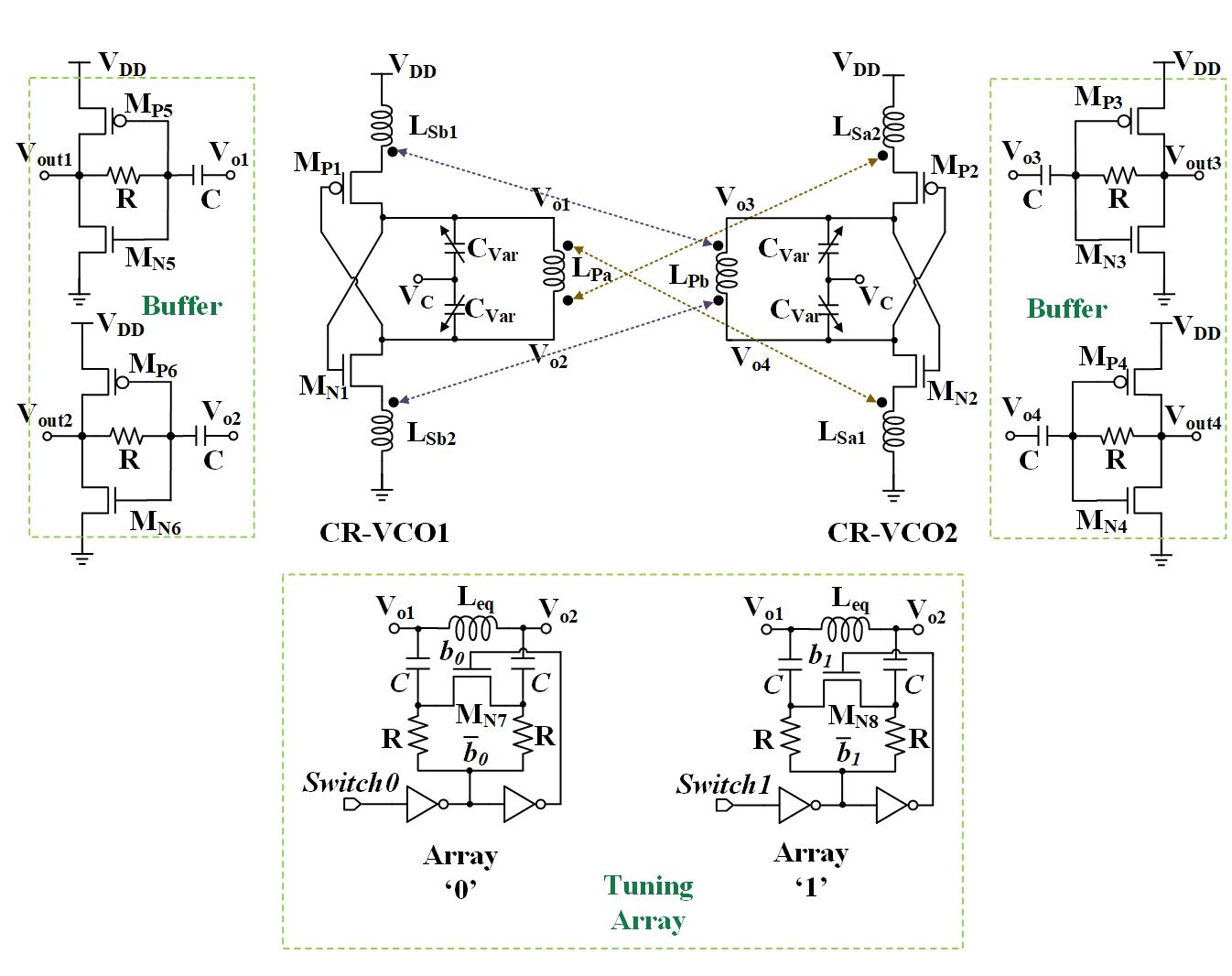}
  \caption{Topology of TC-QVCO with output buffers}
   \label{fig:QuadOsc}
\end{figure*}

\begin{figure*}[!ht] 
  \centering
  \begin{subfigure}[b]{.65\linewidth}
\includegraphics[width=\linewidth]{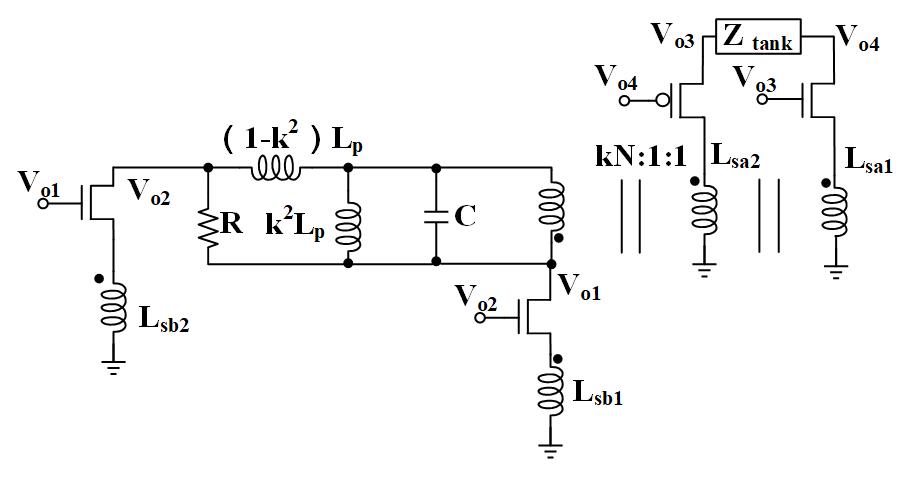}
\caption{Half circuit equivalent small-signal model}\label{fig:withTrans}
\end{subfigure}
\begin{subfigure}[b]{.65\linewidth}
\includegraphics[width=\linewidth]{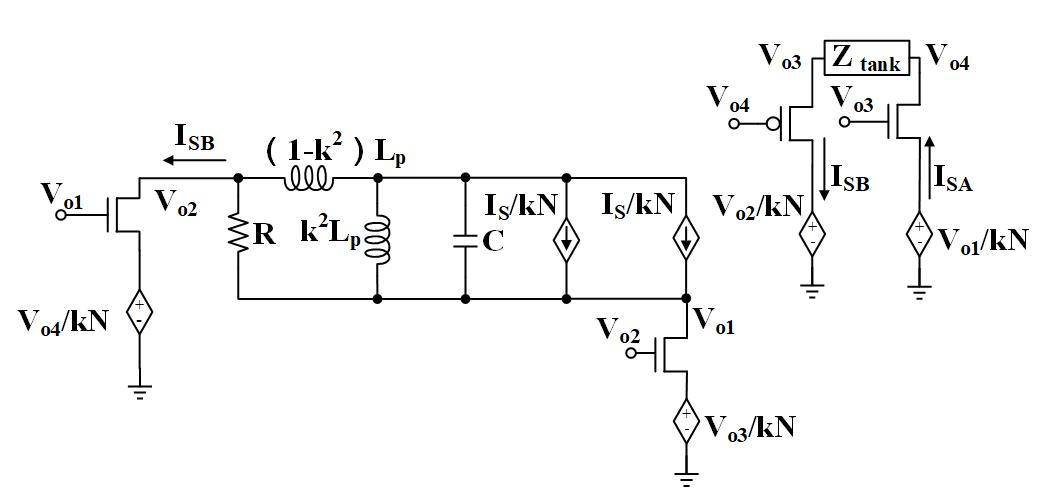}
\caption{Transformers are replaced by controlled source}\label{fig:withoutTrans}
\end{subfigure}
  \caption{Simplified half circuit for QVCO small signal analysis}
   \label{fig:smallSignalHalf}
\end{figure*}

Current reuse VCO (CR-VCO) is another modified circuit of the differential LC-VCO circuit shown in Figure \ref{fig:CROsc}. With this circuit \cite{yun20051mw}, the supply current is lowered by up to 50 \%, which means that the power consumption of this VCO will be reduced to half compared with the conventional differential LC-VCO for the same supply voltage. This supply current reduction is achieved by turning on both PMOS and NMOS of the CR-VCO in the first cycle where the current flows from $V_{dd}$ to $gnd$ through the inductor and in the second cycle both NMOS and PMOS are off and the current flows through the internal capacitances.

As discussed earlier, QVCO with low-power and low metal consumption is highly desired for many applications. In this work, QVCO is designed by combining both transformer-feedback and current-reuse techniques based on \cite{huang20081}. With this approach, power consumption is reduced significantly by reducing supply voltage and supply current but the inductor required to design this QVCO demands metal resources which we will address in Section \ref{sec:TSVIndDes} with the TSV based transformer design.The circuit diagram of the proposed QVCO is shown in Figure \ref{fig:QuadOsc} can be divided into three blocks as shown in Figure \ref{fig:QuadOsc}:
\begin{enumerate} 
\item Main block: It comprises of two identical differential CR-VCOs with two switching transistors for each. Hence, the minimum voltage headroom for this circuit is $2V_{DS}$. Two identical transformers (Transformer A and Transformer B) with cross-coupling between CR-VCOs is used to reduce supply voltage through positive feedback. The primary coil $L_{pa}$ ($L_{pb}$) serves as the inductor in the LC-tank and the secondary coils i.e., $L_{sa1}$ and $L_{sa2}$ ($L_{sb1}$ and $L_{sb2}$) senses the output nodes of the other identical CR-VCO and provide the injection current through cross-coupling. This cross-coupling action ensures that the oscillator outputs maintain quadrature phases.  In addition, the series resistance of the inductors $L_{sa1}$ and $L_{sa2}$ suppresses the amplitude imbalance due to the difference in transconductance ($g_m$)  of NMOS and PMOS transistors, hence no degenerate resistors  are used.

\item Buffers: The outputs from the main block are fed into the buffer circuit because the core of QVCO has no driving ability. This can also isolate the parasitic resistors and capacitors from the LC-tank. These buffers consists of a ac-couple capacitor and an self-bias inverter. Therefore, no additional bias voltage is needed. PMOS and NMOS sizing of these buffers plays a crucial role in obtaining the proper QVCO output  ie., sinusoidal voltages. In our design, we design the buffer to have strong pull-up in order to accommodate the parasitic effects that our QVCO needs to drive.

\item Tuning arrays: Tuning arrays increase the frequency range of the QVCO by varying the center frequency. Tuning array circuit consists of a 2-bit differential binary-weighted swithed-capacitor array (SCAs) as shown in Figure \ref{fig:QuadOsc} where $L_{eq}$ is the equivalent inductance seen between $V_{o1}$ and $V_{o2}$ (The coupling between primary and secondary coils should be considered). The equivalent capacitance to determine center frequency is $C_{Var}$ parallel to $C_{teq}$ where $C_{teq}$ is the equivalent capacitance of the tuning array. The center frequency is varied by varying $C_{teq}$ and voltage controlled tuning is achieved through $C_{Var}$. With this approach, the frequency range of operation of the VCO is increased. Different $C_{teq}$ is obtained by using NMOS switch with a control bit. Let us consider the Array '0' circuit: two capacitors are separated to each other if switch0 is low since the NMOS is OFF else the two capacitors will be in series and accordingly the equivalent capacitance can be 0 or $C/2$. With our circuit, we can have a total of 3 values for $C_{teq}$ because of two arrays and the $C_{teq}$ can be 0,  0.5$C$ and 1.0$C$ when the array control input is `00', `01' (`10') and `11' respectively. We cascade two inverters for each array as shown in the tuning array circuit block to reduce the transient switching times. Also, two large resistors are utilized to isolate the noise from control bits.

\end{enumerate}

To gain more insight on the operation of our QVCO, small signal analysis is performed and the small signal equivalent half circuit is shown in Figure \ref{fig:smallSignalHalf} because of our QVCO symmetry. This half circuit corresponds to the left half of our QVCO and is obtained from the analysis based on \cite{long2017chip}. In this circuit, $R$ and $C$ are equivalent resistance and capacitance of the LC tank parallel to the $k^2 L_p$. The rest $(1-k^2)L_p$ inductance is the self inductance that is not part of the coupling and hence it is connected in series to the equivalent capacitance as shown in the Figure \ref{fig:withTrans}. Also, the rest of the half circuit can be modeled with an ideal transformer and hence can be replaced by controlled sources as shown in Figure \ref{fig:withoutTrans}.

To simplify our analysis, we assume that the transconductances $g_m$ of all transistors to be the same. This assumption is valid since we aim to size the transistors to behave the same way i.e, same transconductances in order to obtain identical outputs in terms of magnitude with desired phase differences between the outputs. With this assumption, the current through the secondary coils i.e., $I_{SA}$ and $I_{SB}$ are same and is assumed to be $I_S$. From Figure \ref{fig:smallSignalHalf}, we can obtain output voltages using Kirchoff's voltage and current laws. The non-coupling primary inductance $(1-k^{2})*L_{p}$ is negligible \cite{cheng20165} and the equation \ref{eq:kirVol} is obtained using Kirchoff's voltage law.

\begin{equation}\label{eq:kirVol}
V_{o2}-V_{o1} + \left[ \left(V_{o1} - \frac{V_{o4}}{kN} \right) g_{m} + \left( \frac{I_S}{kN} + \frac{I_S}{kN} \right) \right] Z_{Tank}  = 0
\end{equation}
where $Z_{tank}$ is given by the equation \ref{eq:impedance} since it is a parallel RLC tank circuit.

\begin{equation} \label{eq:impedance}
    \frac{1}{Z_{tank}}=\frac{1}{R}+\frac{1}{j\omega k^{2}L_p}+j\omega C
\end{equation}

The resonant frequency and the quality factor of this tank circuit are given in equations \ref{eq:resFreq} and \ref{eq:tankQ} respectively.
\begin{equation}\label{eq:resFreq}
    \omega_0 = \frac{1}{\sqrt{k^2 L_p C}}
\end{equation}

\begin{equation}\label{eq:tankQ}
    Q=\omega_0 RC = \frac{R}{\omega_0 k^2 L_p}
\end{equation}
The current $I_S$ can also be expressed using the transistor transconductance as shown in equation \ref{eq:Is}.

\begin{equation}\label{eq:Is}
    I_{S} = \left(V_{o3} - \frac{V_{o1}}{kN} \right)g_{m}
\end{equation}

QVCO produces four quadrature phase oscillations and each CR-VCO in Figure \ref{fig:QuadOsc} produces out-of-phase outputs. Therefore, we can define the relation between the output voltages as shown in eqution \ref{eq:outVolRelation}.

\begin{equation}\label{eq:outVolRelation}
    V_{o1}=-V_{o2}=jV_{o3}=-jV_{o4}
\end{equation}

From equations \ref{eq:kirVol}, \ref{eq:impedance}, \ref{eq:Is} and \ref{eq:outVolRelation}, we obtain equations \ref{eq:gm} and \ref{eq:finalEq2} by equating the real and imaginary parts. The minimum transconductance $g_m$ required for proper oscillation of the QVCO is given in equation \ref{eq:gm}. By solving equation \ref{eq:finalEq2} and replacing $g_m$ from equation \ref{eq:gm}, we obtain the oscillation frequency of the QVCO and is given in the equation \ref{eq:quadFreq}.

\begin{equation}\label{eq:gm}
     g_m=\frac{2}{R}\frac{(kN)^2}{(kN)^2-2}
\end{equation}

\begin{equation}\label{eq:finalEq2}
     \frac{3g_m}{kN}+\frac{2}{\omega{k^2}{L_p}}-2\omega C=0
\end{equation}

\begin{equation}\label{eq:quadFreq}
     \omega = \omega_0 \left[ \frac{3kN}{2Q\left[ \left(kN\right)^2-2 \right]} +\sqrt{ \left(\frac{3kN}{2Q\left[ \left(kN\right)^2-2\right]}\right)^2 +1} \right]
\end{equation}

\section{TSV Inductor Design} \label{sec:TSVIndDes}

In this work, we assume a three-tier 3D process \cite{tida2014novel} for the rest of the paper. The parameters for the TSV inductor of this process is shown in Table \ref{tab:param}. We use a total of 9 metal layers and the thickness of the metal interconnect layer is $30 \mu m$ including the inter-adhesive layer. The TSV inductor is designed such that its cross-section is close to square (TSV spacing equals TSV height) to achieve higher quality factor. The mentioned assumptions are for demonstration purposes only and the TSV inductor design is not limited to these assumptions.

\begin{table}[!ht]
\centering
\caption{Process parameters of the TSV inductor}
\label{tab:param}
\begin{tabular}{|l|l|}
\hline
\textbf{Parameter} 	                        & \textbf{Value} 	\\ \hline
Substrate tier height	                    & 60 $\mu m$	 	\\ \hline
Substrate conductivity ($\sigma$)			&10 $S/m$			\\ \hline
TSV diameter	                            & 20 $\mu m$ 		\\ \hline
Liner thickness	                            & 0.5 $\mu m$		\\ \hline
Minimum TSV pitch                           & 5 $\mu m$         \\ \hline
Inter-adhesive layer thickness		        & 2 $\mu m$		    \\ \hline
Total oxide layer between tiers		        & 30 $\mu m$		\\ \hline
Top metal layer (M9) thickness 			    & 7 $\mu m$		    \\ \hline
Top metal layer (M9) width		            & 24 $\mu m$		\\ \hline
M9-M8 via length 			                & 5 $\mu m$		    \\ \hline
Metal layer (M8) thickness 			        & 7 $\mu m$		    \\ \hline
Metal layer (M8) width		                & 24 $\mu m$		\\ \hline
M8-M7 via length                            & 3 $\mu m$		    \\ \hline  
Metal layer (M7) thickness 			        & 2 $\mu m$		    \\ \hline
Metal layer (M7) width		                & 24 $\mu m$		\\ \hline
\end{tabular}
\end{table}

We implement TSV based 6-port transformer for our QVCO design in two ways based on the TSV inductor models discussed in \cite{tida2014through}: 
\begin{enumerate*}
    \item Toroidal TSV transformer and,
    \item Vertical spiral TSV transformer
\end{enumerate*}. For these transfomers, we need a total of three inductors i.e., one primary and two secondary coils. For our QVCO design shown in Figure \ref{fig:QuadOsc}, we implement two of these transformers that are identical. The TSV transformer structures are detailed in the rest of this section and the specifications with their characteristics are given in Section \ref{subsec:indResults}.

\subsection{Toroidal TSV transformer}

\begin{figure}[!ht] 
  \begin{subfigure}[b]{.95\linewidth}
  \centering
\includegraphics[scale=0.5]{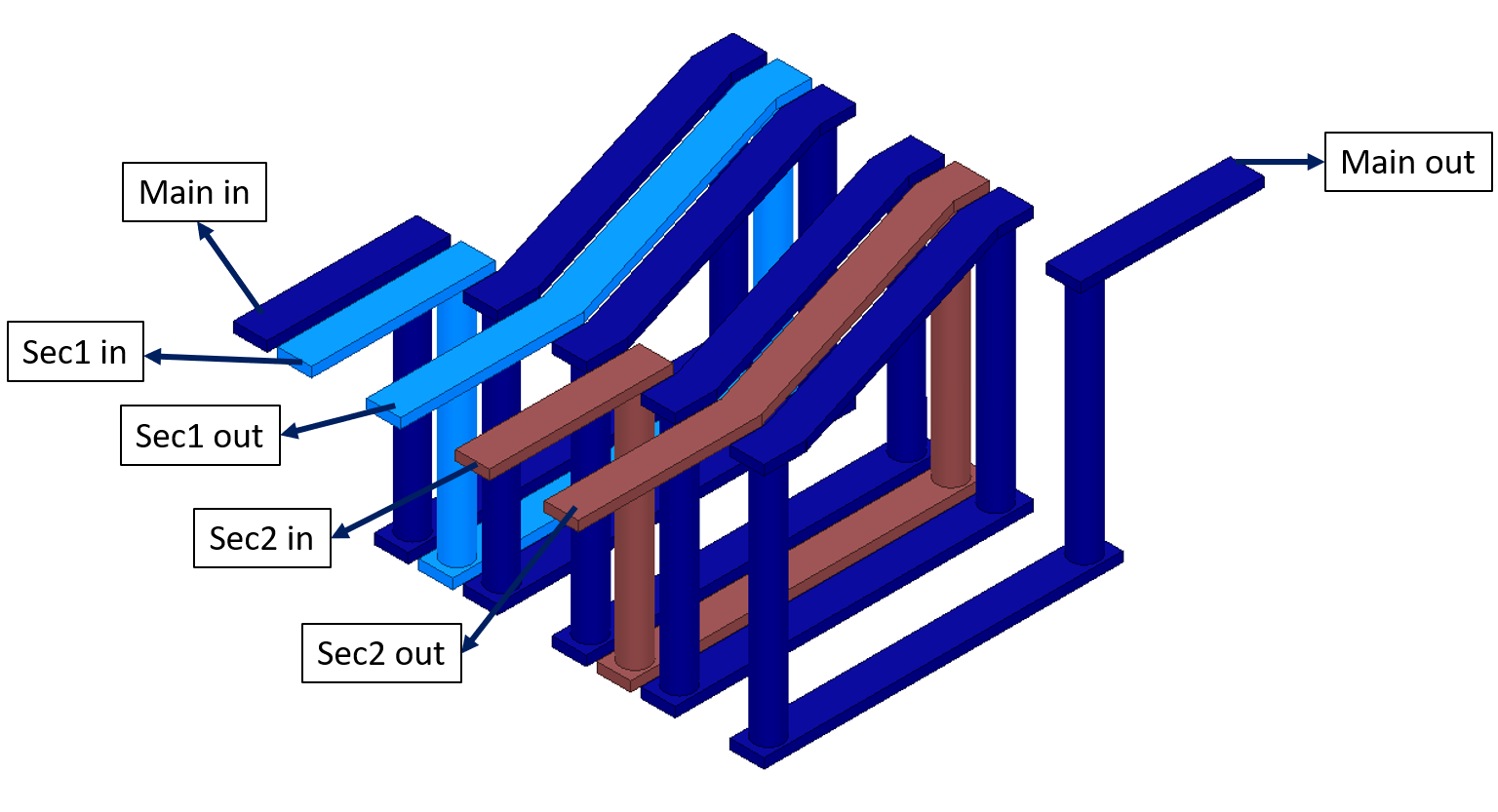}
\caption{cross-view}\label{fig:topview1}
\end{subfigure}
\begin{subfigure}[b]{.9\linewidth}
\centering
\includegraphics[scale=0.5]{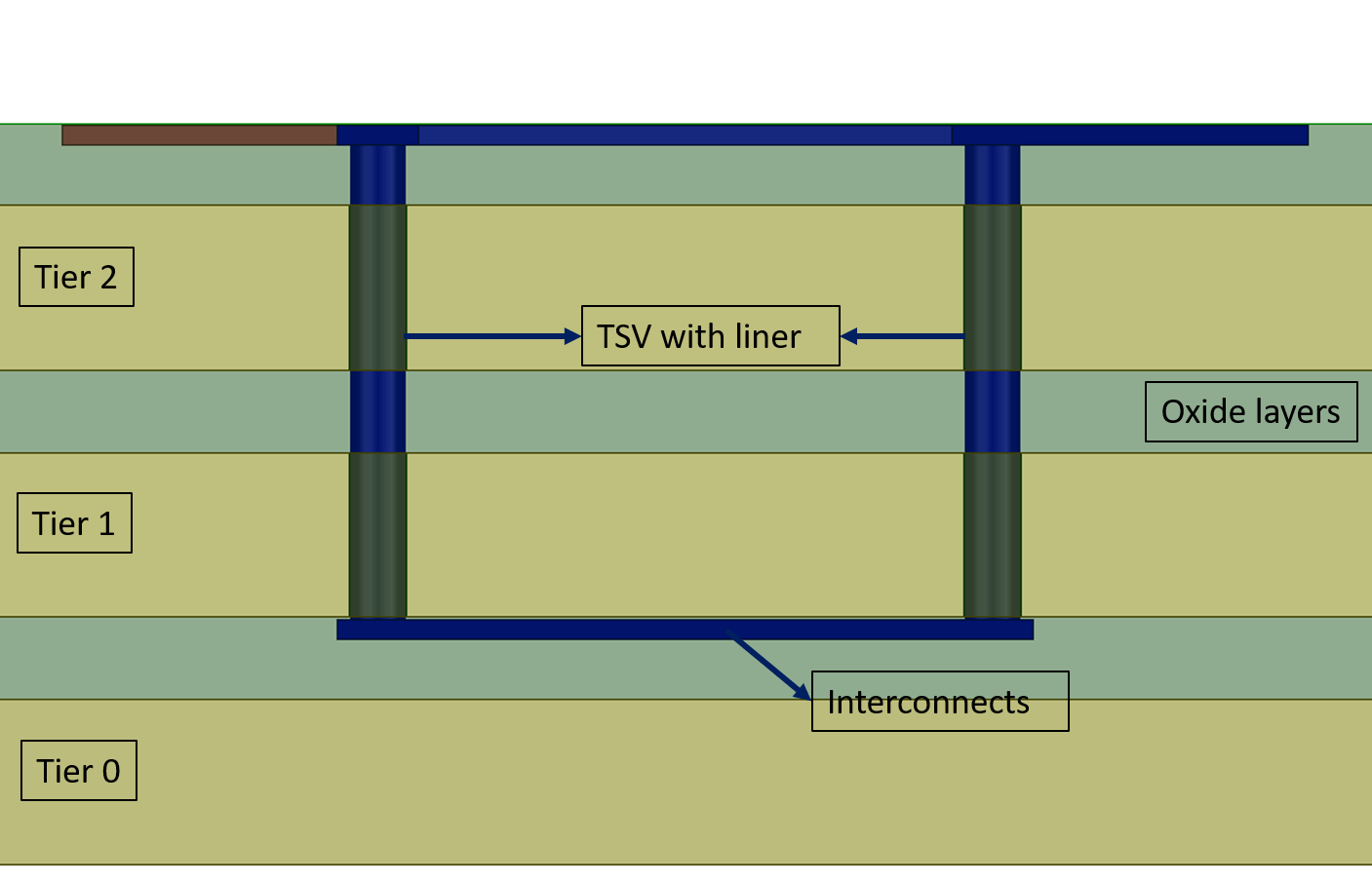}
\caption{side-view}\label{fig:sideview1}
\end{subfigure}
  \caption{Structure of toroidal TSV transformer }
   \label{fig:toroidal}
\end{figure}

On-chip toroidal TSV transformer structure is shown in Figure \ref{fig:toroidal}. TSVs are connected to form a closed loop by using top metal layer and hence the inductor series resistance will be minimum. From the figure, we can see that there are total of three inductors coupled to each other where the primary inductor is named as $ind_1$ and the other two secondary coils are named as $ind_2$ and $ind_3$. For the QVCO design, the primary coil requires higher inductance than the secondary coils. The secondary coils are tightly coupled to the primary coil with the structure shown in Figure \ref{fig:toroidal} where adjacent turns corresponds to different coils. Also, the magnetic coupling between the secondary coils are minimum since they are not tightly coupled.

\begin{figure}[!ht] 
  \begin{subfigure}[b]{.9\linewidth}
  \centering
\includegraphics[scale=0.45]{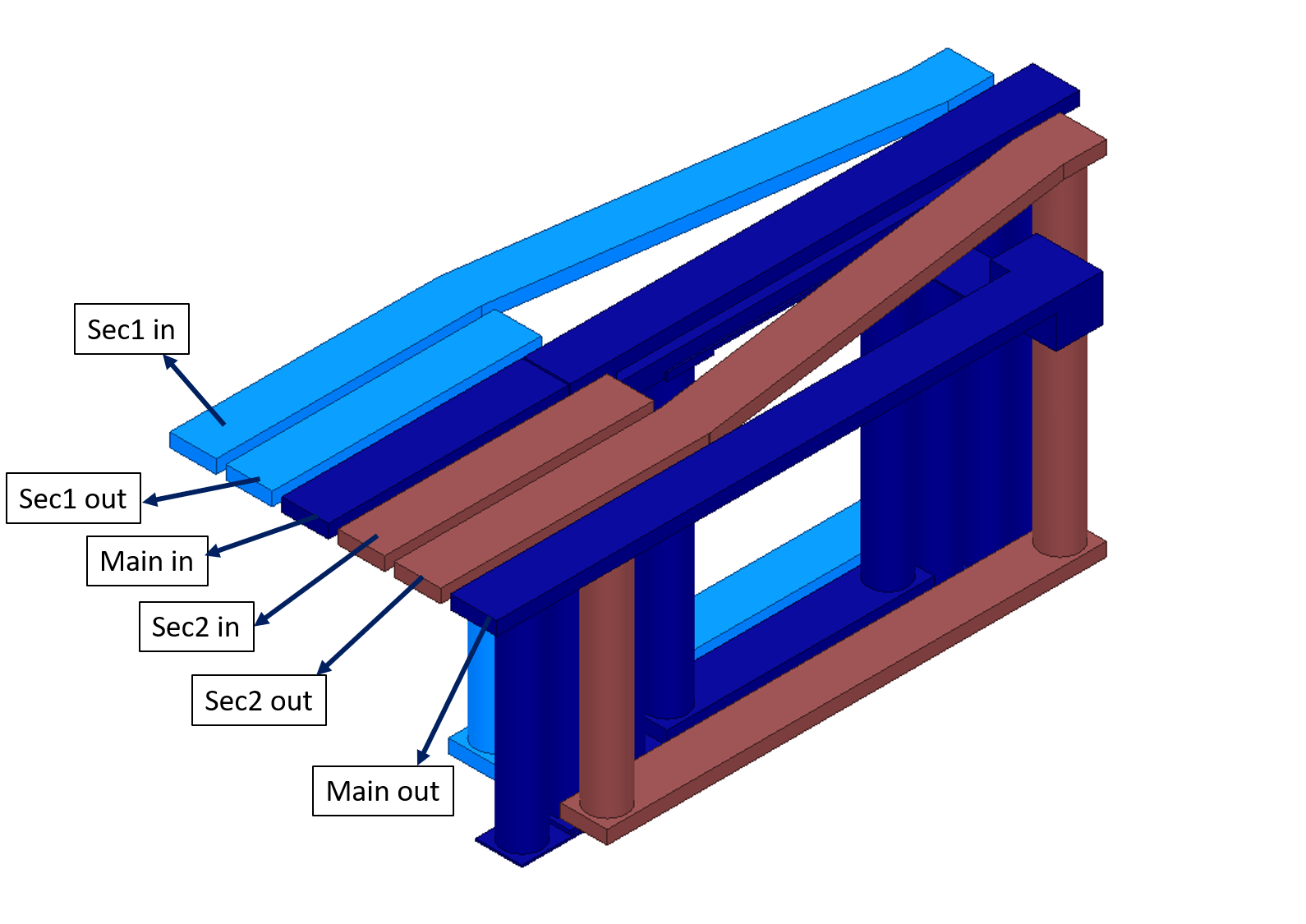}
\caption{cross-view}\label{fig:topview2}
\end{subfigure}
\begin{subfigure}[b]{.9\linewidth}
\centering
\includegraphics[scale=0.45]{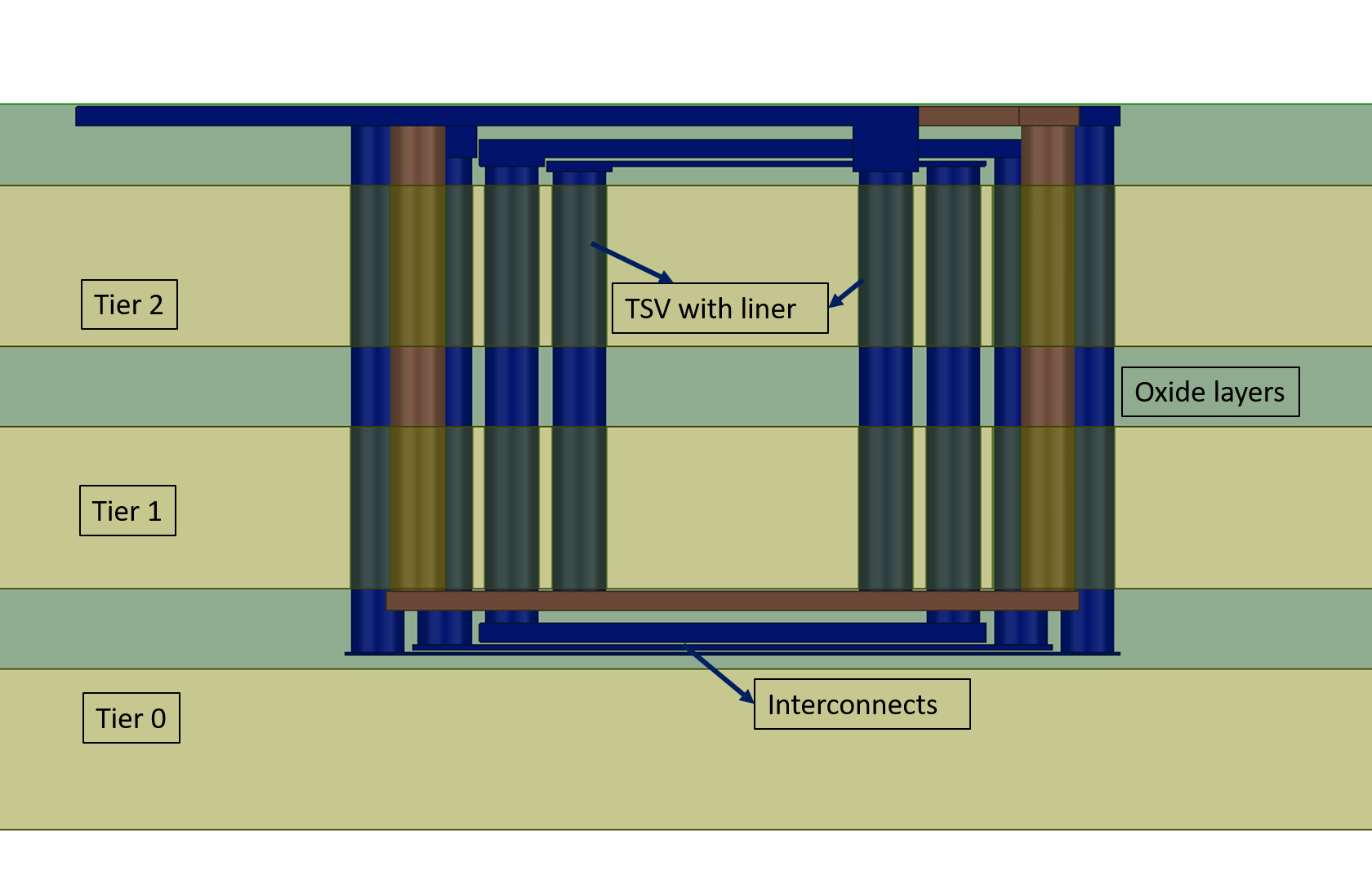}
\caption{side-view}\label{fig:sideview2}
\end{subfigure}
  \caption{Structure of vertical spiral TSV transformer}
   \label{fig:verticalSpiral}
   
\end{figure}

\subsection{Vertical spiral TSV transformer}

Vertical spiral TSV transformer structure is shown in Figure \ref{fig:verticalSpiral}. TSVs are connected to form a closed loop by using top three metal layers and all the TSVs of the same inductor are aligned in a straight line. From the figure, we can see that there are total of three inductors coupled to each other where the primary inductor is named as $Main$ and the other two secondary coils are named as $Sec1$ and $Sec2$. The secondary coils are placed close to the primary coil and the secondary coils are are on the opposite sides to the primary coil as shown in Figure \ref{fig:verticalSpiral}. With this approach, magnetic coupling between  primary coil to second coils  is the possible maximum and between secondary coils is low.

\section{Experimental Results} \label{sec:simResults}

The center frequency of the QVCO is assumed to be $2.5\ GHz$. The capacitance $C_{var}$ ranges from 2.1 $pF$ to 6.3 $pF$ with the control voltage $V_c$ changes from 0.1$V$ to 0.7$V$. Therefore, the primary coil of the transformer should be designed to achieve 3 $nH$ based on Equation \ref{eq:resFreq}, where the $C_{eq}$ is at 4.6 $pF$. To supress the amplitude imbalance and maintain quadrature phases, secondary coils of the transformer is designed to achieve 0.4 $nH$ and k is around 0.52. We implemented the transformer models in commercial full-wave simulation framework and followed the methodology discussed in \cite{tida2014efficacy} to design these inductors based on the specifications obtained.  These inductors are embedded in QVCO design with their s-parameter model with the frequency range from 0 to 20 $GHz$ extracted from full-wave simulator and we use 65 $nm$ TSMC technology library to realize QVCO.  For clarity, we show the design specifications of the QVCO in Table \ref{tab:QVCOSpec}. These design specifications are obtained from industry and are slightly modified.

\begin{table}[!ht]
\centering
\caption{Design Specifications of QVCO}
\label{tab:QVCOSpec}
\begin{tabular}{|l|c|}
\hline
\textbf{Parameter} 	                        & \textbf{Value or Range} 	\\ \hline
Supply voltage $V_{DD}$                     & 0.7 $V$                   \\ \hline    
Center frequency $f_c$	                    & 2.5 $GHz$	 	            \\ \hline
Control voltage $V_c$	            		& 0.1 $V$ to 0.7 $V$			\\ \hline
Primary inductance $L_p$                    & 3 $nH$		            \\ \hline
Secondary inductance $L_s$                  & 0.4 $nH$		            \\ \hline
Capacitance range $C_{var}$                 & 2.1$pF$ to 6.3$pF$      \\ \hline
Rail-to-rail voltage $V_{out_{p-p}}$        & 350 $mV$                  \\ \hline
Max voltage variation $\Delta_{V_{out}}$    & 25 $mV$                   \\ \hline

\end{tabular}
\end{table}

\subsection{TSV inductor design}\label{subsec:indResults}

Design parameters and the performance metrics for toroidal TSV transformer and vertical spiral TSV transformer to implement our QVCO are given in Table \ref{tab:indSpecs}. $L_p$ and $L_s$ represents the primary coil and secondary coils inductance respectively. $R_{pdc}$ and $R_{sdc}$ represents the series DC resistance of primary coil and secondary coils respectively. $R_{pac}$ and $R_{sac}$ represents the series AC resistance of primary coil and secondary coils at frequency $2.5\ GHz$ respectively. $k_{ps}$ and $k_{ss}$ represents the mutual coupling factor between primary coil-secondary coil and between secondary coils respectively. $Area$ represents the total metal resource consumption of the transformer. From the table, we can see that the DC resistance of the primary coil is higher for the vertical spiral transformer compared with the toroidal transformer since thin $M_7$ is used. However, with vertical spiral transformer, the metal resource consumption reduced by up to 18\% compared with the Toroidal transformer. Also, although not shown, we also implemented conventional 2D spiral transformer to compare our TSV transformers and the metal resource consumption reduced by upto 3.2x and 3.9x times by replacing conventional 2D spiral transformer with the toroidal TSV transformer and vertical spiral TSV transformer respectively for the same inductance.
Although not shown in the table, the parasitic capacitance will be higher for vertical spiral TSV based-transformer compared with the toroidal TSV based-transformer due to tight metal interconnects as shown in Figures \ref{fig:toroidal} and \ref{fig:verticalSpiral}.

\begin{table}[!ht]
\centering
\caption{Performance metrics of TSV transformers}
\label{tab:indSpecs}
\begin{tabular}{|c|c|c|}
\hline
\textbf{Parameter} 	                        & \textbf{Toroidal}     & \textbf{Vertical spiral}   	\\ \hline
$L_p$ $(nH)$	                            & 2.99                  & 2.97	 	                    \\ \hline
$R_{pdc}$	$(m\Omega)$              		& 300			        & 488                          \\ \hline
$R_{pac}$	$(\Omega)$              		& 1.4			        & 3.03                            \\ \hline
$L_s$ $(nH)$	                            & 0.38                  & 0.36                   		\\ \hline
$R_{sdc}$	$(m\Omega)$	                    & 64                    & 66 		                    \\ \hline
$R_{sac}$	$(\Omega)$	                    & 0.35                  & 0.38		                    \\ \hline
$k_{ps}$                                    & 0.52                  & 0.54                          \\ \hline
$k_{ss}$		                            & 0.15                  & 0.29		                    \\ \hline
$Area$ $(mm^2)$		                        & 0.17                  & 0.14		                    \\ \hline
\end{tabular}
\end{table}

\subsection{Performance of proposed QVCOs}
We implemented two QVCO designs by using two different TSV transformers discussed in Section \ref{subsec:indResults} i.e., Toroidal TSV based- and, Vertical spiral TSV based- transformers. The performance metrics of the proposed QVCOs are shown in Table \ref{tab:QVCOPer}. The freq range that the proposed VCOs can tune is around 1.4 $GHz$ and the differences in this frequency range between toroidal TSV transformer based QVCO and vertical spiral TSV transformer based QVCO is due to their difference in parasitic capacitances. Since the parasitic capacitance will be higher for vertical spiral TSV based-transformer compared with the toroidal TSV based-transformer  due  to  tight  metal  interconnects, the frequency range is reduced due to the similar transformer inductances. The phase noise $\phi_{noise}$ in the table is at 1 $MHz$ frequency. From the table, we can see that the power consumption of the proposed QVCOs are different. This is because of the difference in series resistances of the transformers. However to sustain oscillation, we need to ensure that the series resistance of primary inductor $R_L$ should be greater than $\frac{2}{g_m}$. The start-up time of the proposed QVCOs i.e., the time required to reach steady state oscillations is around 50$ns$. The common benchmark for LC oscillators is Figure-of-Merit (FoM) \cite{darabi2015radio} and is given by the equation \ref{eq:FoM} and the corresponding FoM values at 1$MHz$ are given in the Table \ref{tab:QVCOPer}.

\begin{equation}\label{eq:FoM}
    FoM = -20\log\Big(\frac{\omega_0}{\Delta \omega}\Big)+10\log(P_{mW})+\phi_{noise}(\Delta \omega)
\end{equation}

\begin{table}[!ht]
\centering
\caption{QVCO Performance metrics}
\label{tab:QVCOPer}
\begin{tabular}{|c|c|c|}
\hline
\textbf{Metric} 	                        & \textbf{Toroidal based-}     & \textbf{Vertical spiral based-}   	\\ \hline
Freq Range ($GHz$)	                            & 2 $\sim$ 3.4                  & 1.83 $\sim$ 3.11	 	                    \\ \hline
$\phi_{noise}$ ($dBc/Hz$)             		& 	-114		        &-111.2                          \\ \hline
$\Delta V_{out}$	($mV$)              		& 19			        & 20                            \\ \hline
Start-up time ($ns$)                               &50               &50                          \\ \hline
Power ($mV$)	                            & 1.5                  & 1.7                   		\\ \hline
FoM (dB)	                            & -180                  & -177                   		\\ \hline
\end{tabular}
\end{table}

\begin{figure}[!ht] 
  \centering
  \includegraphics[scale=0.7]{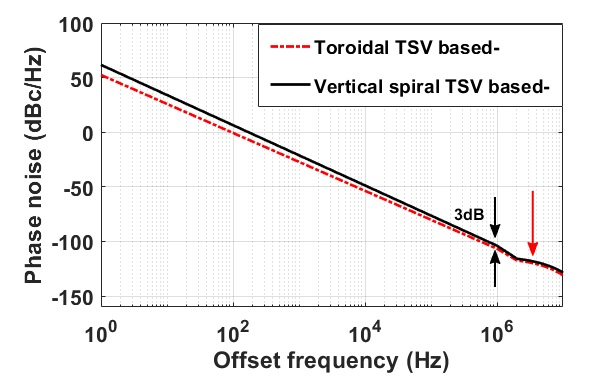}
  \caption{Phase noise v.s. Frequency of proposed QVCOs}
   \label{fig:phiN}
\end{figure}

The phase noise $\phi_{noise}$ vs. offset frequency $f$ is shown in Figure \ref{fig:phiN}. From the figure, we can see that the $\phi_{noise}$ is almost identical for both the QVCO designs since they are different in terms of the transformer performance. $\phi_{noise}$ is higher for vertical spiral TSV transformer based- QVCO compared with the toroidal TSV transformer due to their differences in quality factors  as shown in Table \ref{tab:indSpecs} \cite{hegazi2001filtering}. For example, at 1 $MHz$ frequency offset, the toroidal  TSV based-QVCO has the $\phi_{noise}$ of -114 $dBc/Hz$, almost 3$dB$ lower compared with the vertical TSV based-QVCO. 

\begin{figure}[!ht] 
  \centering
  \includegraphics[scale=0.7]{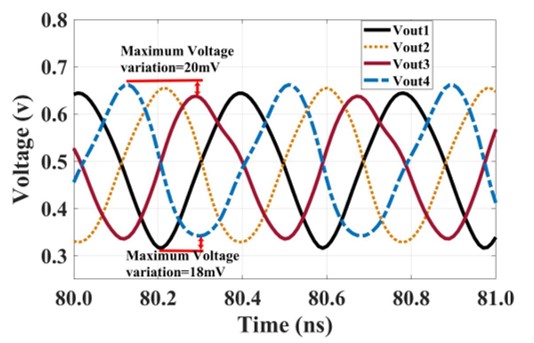}
  \caption{Output voltage of vertical sprial transformer based- QVCO}
   \label{fig:output}
\end{figure}

The steady state quadrature output voltages of vertical sprial transformer based-QVCO is shown in Figure \ref{fig:output}. The frequency of these output voltages are 2.5 $GHz$ and the rail to rail voltage is around 350 $mV$ for a 0.7 $V$ supply. From the figure, we can see that the output voltages does not have same rail to rail voltage and this difference is due to the variations in transconductances of NMOS and PMOS transistors and in drain-source voltage of the transistors. However, the maximum voltage difference between the outputs is less than 20 $mV$ and is well with in the limits of the design specifications.

\begin{table*}[!ht]
\centering
\caption{Comparison with the state-of-the arts}
\label{tab:comparison}

\scalebox{0.79}{\begin{tabular}{||c c c c c c c c||}
\hline
\textbf{Specification} 	& \textbf{Vertical Spiral TSV}  & \textbf{Toroidal TSV}  & \textbf{\cite{huang2014ultra}}                  &\textbf{\cite{yahalom2015vertical}}           &\textbf{\cite{ding2018low}}                  &\textbf{\cite{wang2012low}}                  &\textbf{\cite{wang2014frequency}}	\\ \hline
Technology	            &65nm                          &65nm	 	                  &90nm             
&28nm                   &40nm                          &65nm                          &180nm            \\ \hline
CMOS                    &3D                            &3D                            &3D       
&3D                      &-                             &-                            &-                \\ \hline
Freq. Range ($GHz$)       &1.83 $\sim$ 3.11              &2 $\sim$ 3.4                  &43.3 $\sim$ 43.9     
&1.67 $\sim$ 2.09       & 1.55 $\sim$ 1.67                       &1.35 $\sim$ 1.75              &3.17 $\sim$ 5.27  \\ \hline   
$\phi_{noise}$ at $1MHz(dBc/Hz)$ &-111.2                       &-114                          &-90.83       
&-122                    &-118	                       &-120	                      &-115              \\ \hline
Supply Voltage ($V$)	     &0.7	                           &0.7	                          &1.2	
&1	                     &0.6	                       &1	                          &0.65              \\ \hline
Power ($mW$)	             &1.7	                       &1.5	                          &18.5	
&5	                     &1.32	                       &2.6	                          &2.37              \\ \hline
FOM at 1$MHz$ ($dB$)	         &-177	                       &-180	                      &-171	
&-181	                 &-184.4	                   &-180	                      &-185             \\ \hline
Die Area ($mm^2$)	     &0.13	                       &0.19	                      &0.2	                        &Unknown	             &0.62	                       &0.28	                      &0.828            \\ \hline
\end{tabular}}
\end{table*}

\subsection{Comparison of proposed QVCO designs with other works}
Table \ref{tab:comparison} summarizes the comparison of proposed QVCOs with other current QVCO designs. From the table, we can see that the proposed QVCO performance is on par with the other current QVCO designs. In terms of power consumption, proposed QVCOs are superior to other works except with the QVCO design in \cite{ding2018low} but the die area and the frequency range is higher for the proposed design. In terms of die area, proposed QVCO design has the minimum die area and this is achieved through the proposed TSV transformer structure utilization to design QVCOs. In addtion, proposed QVCO designs  demonstrate relatively high phase noise. Although \cite{huang2014ultra, yahalom2015vertical, ding2018low} used 3D solenoid inductor to implement QVCOs, they use a single conventional TSV inductor based QVCOs instead of taking advantage of coupling mechanisms of the transformer designs.

\section{Conclusion}\label{sec:conclusions}
In this article, we have demonstrated the effectiveness of TSV transformer structures in the QVCO designs in 3D ICs. We also proposed a novel QVCO that can be used to provide higher frequency range. Experimental results show  by replacing conventional spiral transformers with TSV transformers, up to 3.9x reduction in metal resource consumption. The proposed QVCOs achieves a phase noise of $-114$  $dBc/Hz$@1 $MHz$ and $-111.2$ $dBc/Hz$@1 $MHz$ at the carrier of 2.5 $GHz$ for toroidal TSV transformed based-QVCO and vertical spiral transformer based-QVCO respectively. The power consumption is only 1.5 $mW$ and 1.7 $mW$ for toroidal TSV transformed based-QVCO and vertical spiral transformer based-QVCO respectively, under the supply voltage of 0.7 $V$. We have also discussed the comparison of our works with the current QVCO designs.


%



\section*{Acknowledgment}
This work is partially supported by the National Natural Science Foundation of China (Grant No. 61801524).






\bibliographystyle{IEEEtran}
\bibliography{uma_bibtex}

%




\end{document}